\documentclass[aps,twocolumn,pra,superscriptaddress,amsmath,amssymb]{revtex4-1}
\usepackage{dcolumn}
\usepackage{bm}
\usepackage{amsmath}
\usepackage{txfonts}
\usepackage[T1]{fontenc}
\usepackage{xspace}
\usepackage{ulem}
\usepackage{braket}
\ifx\pdfoutput\undefined
\usepackage[dvipdfmx]{graphicx}
\usepackage[dvipdfmx]{hyperref}
\usepackage[dvipdfmx]{color}
\else
\usepackage{graphicx}
\usepackage{hyperref}
\usepackage{color}
\fi

\begin{document}
\let\emph\textit

\title{
Majorana correlations in the Kitaev model with ordered-flux structures
}

\author{Akihisa Koga}

\author{Yuta Murakami}

\affiliation{
  Department of Physics, Tokyo Institute of Technology,
  Meguro, Tokyo 152-8551, Japan
}

\author{Joji Nasu}

\affiliation{
  Department of Physics, Yokohama National University, Hodogaya, Yokohama 240-8501, Japan
}
\affiliation{
  PRESTO, Japan Science and Technology Agency, Honcho Kawaguchi, Saitama 332-0012, Japan
}

\date{\today}
\begin{abstract}
  We study the effects of the flux configurations on the emergent Majorana fermions 
  in the $S=1/2$ Kitaev model on 
  a honeycomb lattice, where quantum spins are fractionalized
  into itinerant Majorana fermions and localized fluxes.
  A quantum spin liquid appears as the ground state of the Kitaev model in the flux-free sector,
  which has intensively been investigated so far.
  In this flux sector, the Majorana fermion system
  has linear dispersions and shows 
  power law behavior in the Majorana correlations.
  On the other hand, periodically-arranged flux configurations
  yield low-energy excitations in the Majorana fermion system,
  which are distinctly different from those in the flux-free state.
  We find that one of the periodically arranged flux states results in the
  gapped Majorana dispersion and the exponential decay  
  in the Majorana correlations.
  The Kitaev system with another flux configuration exhibits
  a semi-Dirac like dispersion,
  leading to the power law decay with
  a smaller power than that in the flux-free sector
  along symmetry axes.
  We also examine the effect of the randomness in the flux configurations
  and clarify that the Majorana density of states is filled by increasing the flux density,
  and power-law decay in the Majorana correlations remains.
  The present results could be important to control the motion of Majorana fermions,
  which carries the spin excitations, in the Kitaev candidate materials.
\end{abstract}
\maketitle

\section{Introduction}
Spin transport in the quantum spin systems has attracted much interest as a fundamental magnetic phenomenon but also in applications to spintronics.
In the insulating magnets,
the spin degrees of freedom are carried by the magnons
in the magnetically ordered states with long-range spin-spin
correlations.
A flow of the spin angular momentum has been observed experimentally
in the compounds such as $\rm Y_3Fe_5O_{12}$ and
$\rm LaY_2Fe_5O_{12}$~\cite{uchida2010,uchida2010spin,Xiao2010,Rezende2014}.
By contrast, the spin transport in the nonmagnetic state has been discussed recently.
One of the typical examples is the one-dimensional antiferromagnetic Heisenberg
chain~\cite{PhysRevB.88.041405,hirobe2017one},
where a quantum spin liquid is realized with quasi-long range spin-spin correlations.
The measurement of the spin Seebeck effect in the candidate compound $\rm Sr_2CuO_3$
has clarified that
the spinons can carry the excitations of the spin degree of freedom~\cite{hirobe2017one}.
In this system, the spin transport is governed by the gapless dispersion of
the spinons originating from the quasi-long range spin-spin correlations.

Another candidate of quantum spin liquids is provided by
the Kitaev quantum spin model~\cite{Kitaev2006},
which has  been extensively examined~\cite{PhysRevLett.100.177204,Jackeli_2009,Chaloupka_2010,PhysRevLett.117.277202,Yamaji_2014,Knolle_2014,Nasu1,Nasu2,Suzuki_2015,RevJPSJ}.
This model is exactly solvable and possesses two types of elementary excitations:
itinerant Majorana fermions and localized fluxes~\cite{Kitaev2006}.
In the ground-state flux configuration, the Majorana fermion system is gapless
while spin correlations are extremely short-ranged, and a spin gap exists in the magnetic excitations.
In our previous paper~\cite{Minakawa}, we have examined the spin transport in the Kitaev model.
We have found that 
the spin excitation propagates in the Kitaev quantum spin liquid
despite the short-ranged spin correlations.
Also, it has been clarified that the spin transport is carried by the Majorana fermions and its velocity
corresponds to the slope of the 
linear dispersion~\cite{Taguchi}.
These suggest the importance of the Majorana correlations 
for
the spin transport in the Kitaev system.
Nevertheless, these correlations have not been discussed in detail.
Furthermore, it is not clear how Majorana correlations are affected
by the spatial distribution of fluxes,
which are the other degrees of freedom owing to the spin fractionalization.
Then, a question arises: Is it possible to manipulate the motion of
itinerant Majorana fermions
and to generate a ``Majorana insulator'' in terms of the fluxes?
It is highly desired to 
examine how the spin transport mediated by the Majorana fermions is controlled by the flux degree of freedom 
when one considers the realistic spintronic devices using the Kitaev candidate materials
such as $\rm A_2IrO_3$ (A=Li, Na, Cu)\cite{PhysRevB.82.064412,PhysRevLett.108.127203,PhysRevLett.109.266406,PhysRevLett.108.127204,Kitagawa2018nature,PhysRevLett.114.077202},
$\alpha$-$\rm RuCl_3$~\cite{Plumb2014,Kubota2015},
$\rm YbCl_3$\cite{PhysRevB.102.014427}, and $\rm Os_{\it x}Cl_3$~\cite{Kataoka}.
This problem should be common to the honeycomb lattice systems with linear dispersions
like graphene, where the periodic defects or impurities make
the Dirac semi-metallic system insulating~\cite{Fujimoto,Yamagami}.

Motivated by this,
we investigate how the Majorana correlations are affected by the flux configurations
in the Kitaev model.
In particular, we focus on two uniform flux and two ordered-flux configurations.
We find that the linear dispersions are present in the uniform cases and
the Majorana correlations are well scaled by its velocity and the number of the point node.
We also find that the periodically-aligned flux configurations induce distinct low-energy features, gapped and semi-Dirac type dispersions.
The present results suggest that the flux configurations play a crucial role
for the low-energy properties of the Majorana fermions.
Since these mediate the spin transport in the Kitaev spin liquids,
we expect that the mobility can be manipulated by the flux degree of freedom,
which might be useful for applications to spintronics devices.
We also address how the random flux configuration affects
Majorana correlations,
which may be important to discuss the stability of the spin transport
against thermal fluctuations.

The paper is organized as follows.
In. Sec.~\ref{model},
we introduce the Kitaev model on the honeycomb lattice
and briefly explain our methods.
The low-energy properties and Majorana correlations
for the systems with flux configurations are
discussed in Sec.~\ref{result}.
A summary is given in the last section.
The effects of the three-spin interactions
are discussed in Appendix A.

\section{Model and Hamiltonian}\label{model}
\begin{figure}[htb]
 \includegraphics[width=\linewidth]{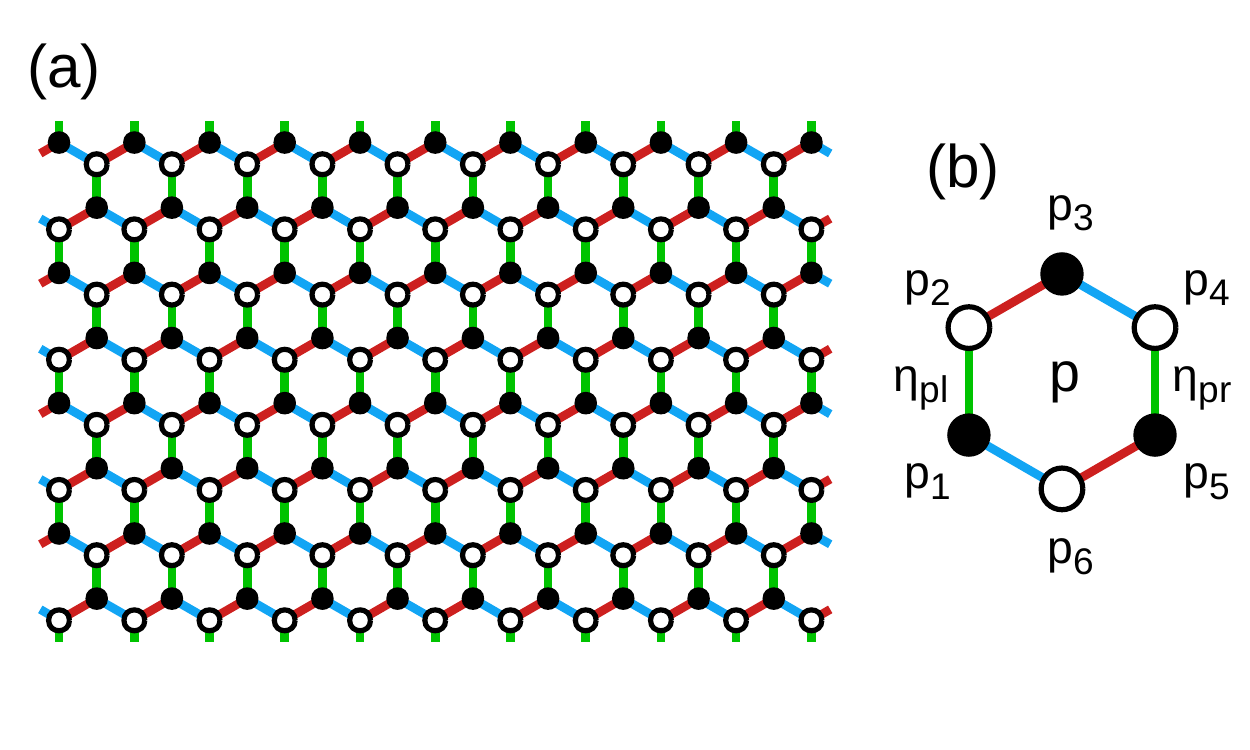}
 \caption{
   (a) $S=1/2$ Kitaev model on the honeycomb lattice.
   Red, blue, and green lines represent the $x$-, $y$-, and $z$-bonds,
   respectively.
   (b) Hexagonal plaquette with sites 
    $p_1$, $p_2$, $\cdots$, $p_6$
   shown for the local operator $W_p$.
   $\eta_{pl}$ and $\eta_{pr}$ denote the left and right $z$-bonds on the plaquette.
 }
 \label{simple}
\end{figure}

We consider the Kitaev model on the honeycomb lattice given by
\begin{eqnarray}
  H&=&-J_x\sum_{\langle i,j \rangle_x}S_i^x S_j^x
  -J_y\sum_{\langle i,j \rangle_y}S_i^y S_j^y
  -J_z\sum_{\langle i,j \rangle_z}S_i^z S_j^z,
\label{eq:Model}
\end{eqnarray}
where
$\langle i,j\rangle_\alpha$ stands for the nearest-neighbor pair
on the $\alpha(=x,y,z)$ bonds, as depicted in Fig.~\ref{simple}(a).
$S_i^\alpha(=\frac{1}{2} \sigma_i^\alpha)$
is the $\alpha$ component of the $S=1/2$ spin at $i$th site
and $\sigma^\alpha$ is the $\alpha$ component of the Pauli matrices.
$J_\alpha$ is the exchange coupling on the $\alpha$ bonds.
In this model, there are local conserved quantities on hexagonal plaquettes in the honeycomb lattice.
The corresponding operator $W_p$ is defined on each plaquette $p$ as
\begin{eqnarray}
  W_p&=&\sigma_{p_1}^x\sigma_{p_2}^y
  \sigma_{p_3}^z\sigma_{p_4}^x\sigma_{p_5}^y\sigma_{p_6}^z, \label{wp}
\end{eqnarray}
where $p_i(i=1,2,\cdots, 6)$ is the site in the plaquette $p$
[see Fig.~\ref{simple}(b)].
One finds $W_p^2=1$, $[H, W_p]=0$ and $[W_p, W_q]=0$.
Therefore, the eigenstates for the Hamiltonian Eq.~(\ref{eq:Model}) can be
specified by the set of $w_p$, $|\psi\rangle=|\psi;\{w_p\}\rangle$,
where $w_p(=\pm 1)$ is the eigenvalue of the local operator $W_p$.
The anticommutation relation $\{W_p, S_{p_1}^y\}=0$ leads to
the absence of the spin moment $(\langle \psi|S_{p_1}^y|\psi\rangle=0)$,
and long-range spin-spin correlations
$[\langle \psi|S_{p_1}^y S_i^\alpha|\psi\rangle=0\;
  (i\neq p_6$ and $\alpha\neq y)]$.
Since this relation is satisfied for
each topologically equivalent plaquette in the system,
the quantum spin liquid state without spin-spin correlations beyond nearest neighbor sites
is realized for any configuration of $w_p$.
It is known that the ground state belongs to the subspace with $w_p=1$
for all plaquettes~\cite{Kitaev2006}.
This allows us to regard a plaquette with $w_p=-1$ as an excited flux and the subspace for the ground state can be identified
as a flux-free sector.

To discuss how flux configurations affect low-energy properties,
we introduce a Majorana representation for the Kitaev spin model given in Eq~\eqref{eq:Model}. 
First, we make use of Jordan-Wigner transformations
$S_i^+=\prod_{i'}^{i-1}\left(1-2n_{i'}\right)a_i^\dag,
S_i^-=\prod_{i'}^{i-1}\left(1-2n_{i'}\right)a_i,
S_i^z=n_i-\frac{1}{2}$,
where $a_i^\dag$ and $a_i$ are the creation and annihilation operator
of the fermion at $i$th site.
The Hamiltonian is then expressed as
\begin{eqnarray}
  H&=&-\frac{J_x}{4}\sum_{(rb,r'w)_x}
  \left(a_{rb}-a_{rw}^\dag\right)\left(a_{r'w}+a_{r'w}^\dag\right)\nonumber\\
  &&-\frac{J_y}{4}\sum_{(rb,r'w)_y}
  \left(a_{rb}+a_{rw}^\dag\right)\left(a_{r'w}-a_{r'w}^\dag\right)\nonumber\\
  &&-\frac{J_z}{4}\sum_r
  \left(2n_{rb}-1\right)\left(2n_{rw}-1\right),
\end{eqnarray}
where $a_{rb} (a_{rw})$ is an annihilation operator of the fermion
at the black (white) site on the $r$th $z$-bond and 
$(rb,r'w)_\alpha$ means the nearest neighbor pair linked
by the $\alpha$-bond [see Fig.~\ref{simple}(a)].
Furthermore, Majorana fermion operators $\gamma, \bar{\gamma}$ are
introduced~\cite{Chen_2007,Feng_2007,Chen_2008} as,
\begin{equation}
  \left\{
  \begin{array}{rcl}
    i\gamma_{rw}&=&a_{rw}-a_{rw}^\dag\\
    \bar{\gamma}_{rw}&=&a_{rw}+a_{rw}^\dag    
  \end{array}
  \right.,\hspace{5mm}
  \left\{
  \begin{array}{rcl}
    \gamma_{rb}&=&a_{rb}+a_{rb}^\dag\\
    i\bar{\gamma}_{rb}&=&a_{rb}-a_{rb}^\dag
  \end{array}
  \right.,
\end{equation}
where Majorana operators satisfy
$\gamma_i^\dag=\gamma_i$, $\bar{\gamma}_i^\dag=\bar{\gamma}_i$,
$\{\gamma_i,\gamma_j\}=\{\bar{\gamma}_i,\bar{\gamma}_j\}=2\delta_{ij}$.
The Hamiltonian Eq.~(\ref{eq:Model}) is then rewritten as,
\begin{eqnarray}
  H=
  -\frac{iJ_x}{4}\sum_{(rb,r'w)_x}\gamma_{rb}\gamma_{r'w}
  -\frac{iJ_y}{4}\sum_{(rb,r'w)_y}\gamma_{rb}\gamma_{r'w}
  -\frac{iJ_z}{4}\sum_r \eta_r\gamma_{rb}\gamma_{rw},\nonumber\\
  \label{eq:Majorana}
\end{eqnarray}
where $\eta_r=i\bar{\gamma}_{rb}\bar{\gamma}_{rw}$.
Since $[H,\eta_r]=0, [\eta_r, \eta_{r'}]=0$, and $\eta_r^2=1$,
$\eta_r$ is a $Z_2$ local conserved quantity.
Namely, on a certain plaquette, the local operator $W_p$ is represented as $W_p=\eta_{pl} \eta_{pr}$,
where $\eta_{pl}$ and $\eta_{pr}$
are defined on the left and right $z$-bonds on the plaquette $p$
[see Fig.~\ref{simple}(b)].
Then, each flux configuration is described by the set of $\{\eta_r\}$
and low-energy properties can be discussed on the basis of Eq.~(\ref{eq:Majorana}).

\begin{figure}[htb]
 \includegraphics[width=\linewidth]{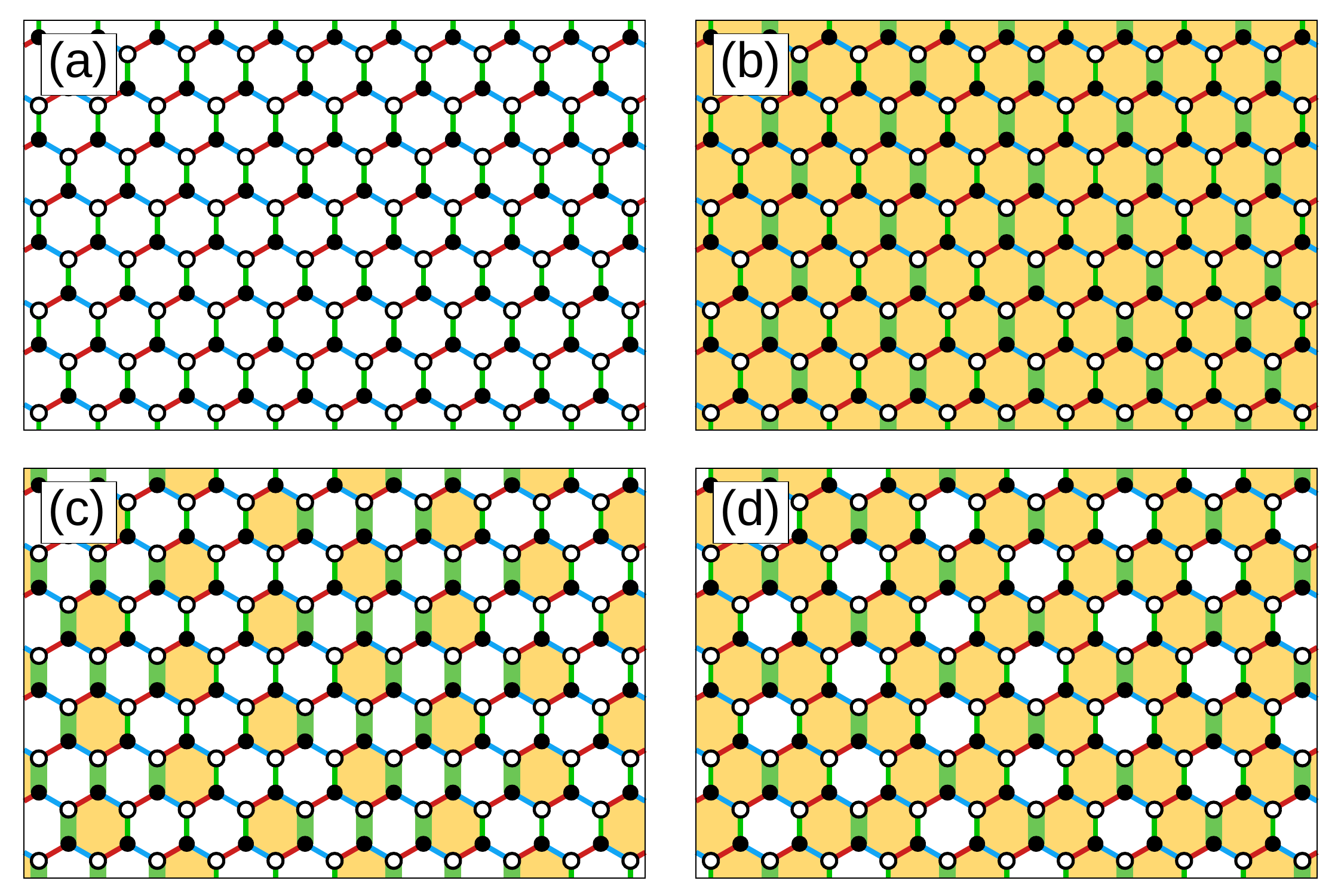}
 \caption{
   Four flux configurations in the Kitaev model on the honeycomb lattice.
   Each empty (shaded) hexagon on the plaquette $p$
   represents the eigenvalue of the local conserved quantity
   $w_p=1 (w_p=-1)$.
   Thin (bold) green lines on the $r$th $z$-bond
   represent $\eta_r=1$ ($\eta_r=-1$).
   The ground state (the configuration I) of the Kitaev model belongs to the subspace
   with $\{w_p=1\}$ (a).
   The configuration II is specified by the full-flux state (b).
   (c) and (d) represent the ordered-flux configurations III and IV, respectively.
 }
 \label{lattice}
\end{figure}

In the present study, 
we focus on two uniform flux configurations
and two periodically-aligned flux configurations, as shown in Fig.~\ref{lattice}.
Diagonalizing the Majorana Hamiltonian in each flux configuration, 
we calculate the dispersion relation of
the itinerant Majorana fermions $E({\bf k})$ and
its density of states $\rho(E)$.
Furthermore, we calculate Majorana correlation functions
$C(d_{rr'})=|\langle \gamma_{rb} \gamma_{r'w}\rangle|$
along the horizontal axis as one of symmetry axes, where $d_{rr'}=|r-r'|$.
Namely, Majorana correlations in the same sublattice are exactly zero
$\langle \gamma_{rb}\gamma_{r'b}\rangle =\langle \gamma_{rw}\gamma_{r'w}\rangle =0$
for arbitrary $r$ and $r'$ since the system is bipartite.

\begin{figure}[htb]
 \centering
 \includegraphics[width=\linewidth]{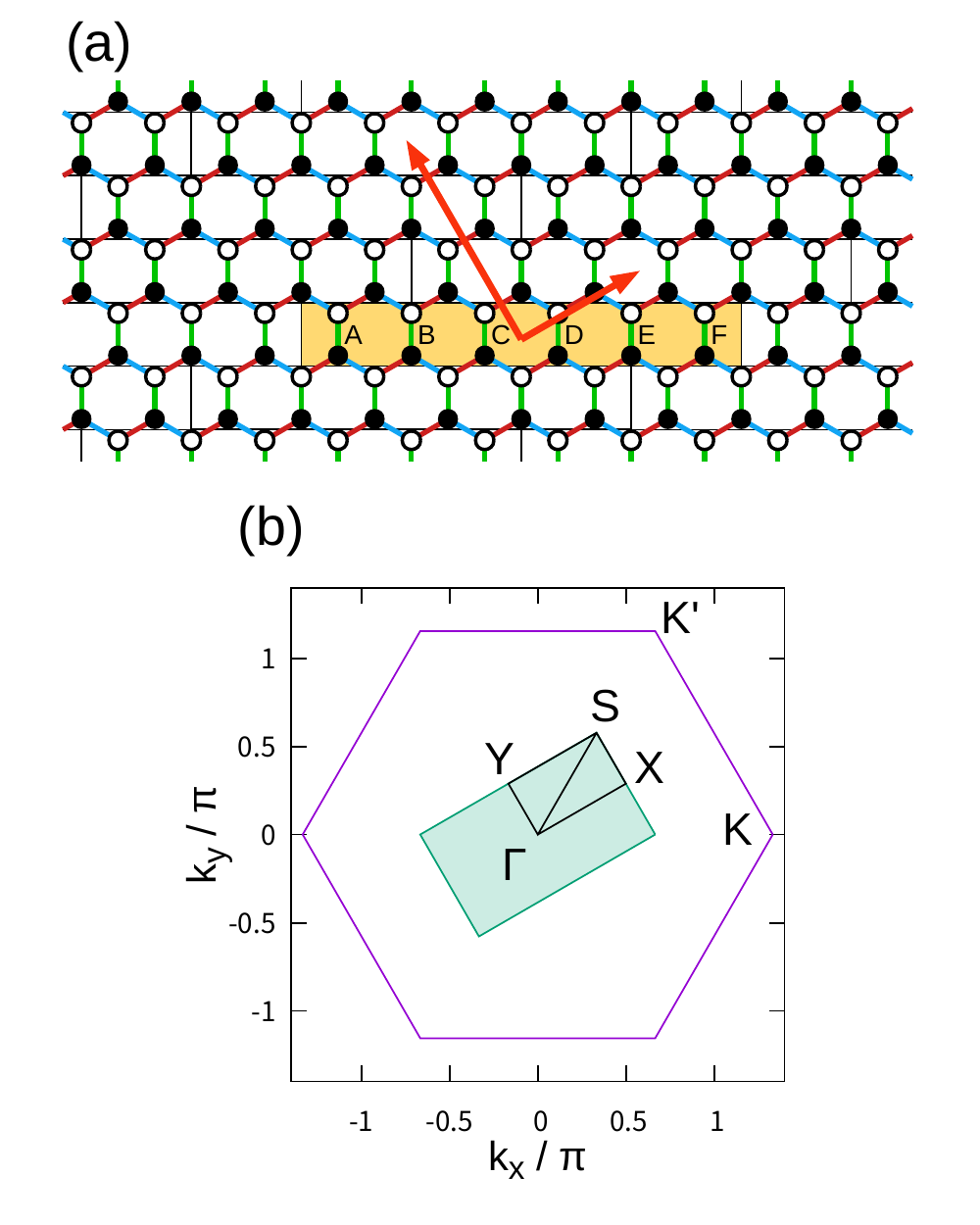}
 \caption{
   (a) Shaded region represents the unit cell and
   two bold arrows represent the translation vectors for the unit cell.
   (b) The corresponding Brillouin zone.
 }
 \label{fig:BZ}
\end{figure}

Here, we introduce the unit cell including six $z$-bonds
with $\{\eta_A,\eta_B,\cdots,\eta_F\}$,
as shown in Fig.~\ref{fig:BZ}(a). 
This allows us to treat four configurations shown in Fig.~\ref{lattice} on an equal footing
for the Majorana representation given in Eq.~(\ref{eq:Majorana}).
The uniform flux configurations I, II and ordered-flux configurations III, IV are
specified by $\eta^{({\rm I})}=\{1,1,1,1,1,1\}$, $\eta^{({\rm II})}=\{1,-1,1,-1,1,-1\}$,
$\eta^{({\rm III)}}=\{1,1,1,-1,-1,-1\}$, and
$\eta^{({\rm IV})}=\{1,-1,1,1,-1,1\}$, respectively.
The Brillouin zone for the Kitaev model is originally given by the hexagon
in Fig.~\ref{fig:BZ}(b), where the lattice constant is set as a unit of length.
The reduced one for the large unit cell is represented by the rectangle,
which is shown as the shaded area.
Hereafter, we restrict our discussions
to the isotropic case with $J=J_x=J_y=J_z$.
We then examine the low-energy excitations in the Kitaev models
with the different flux configurations.

\section{Results}\label{result}

\subsection{Ordered-flux configurations}
\begin{figure}[htb]
 \centering
 \includegraphics[width=\linewidth]{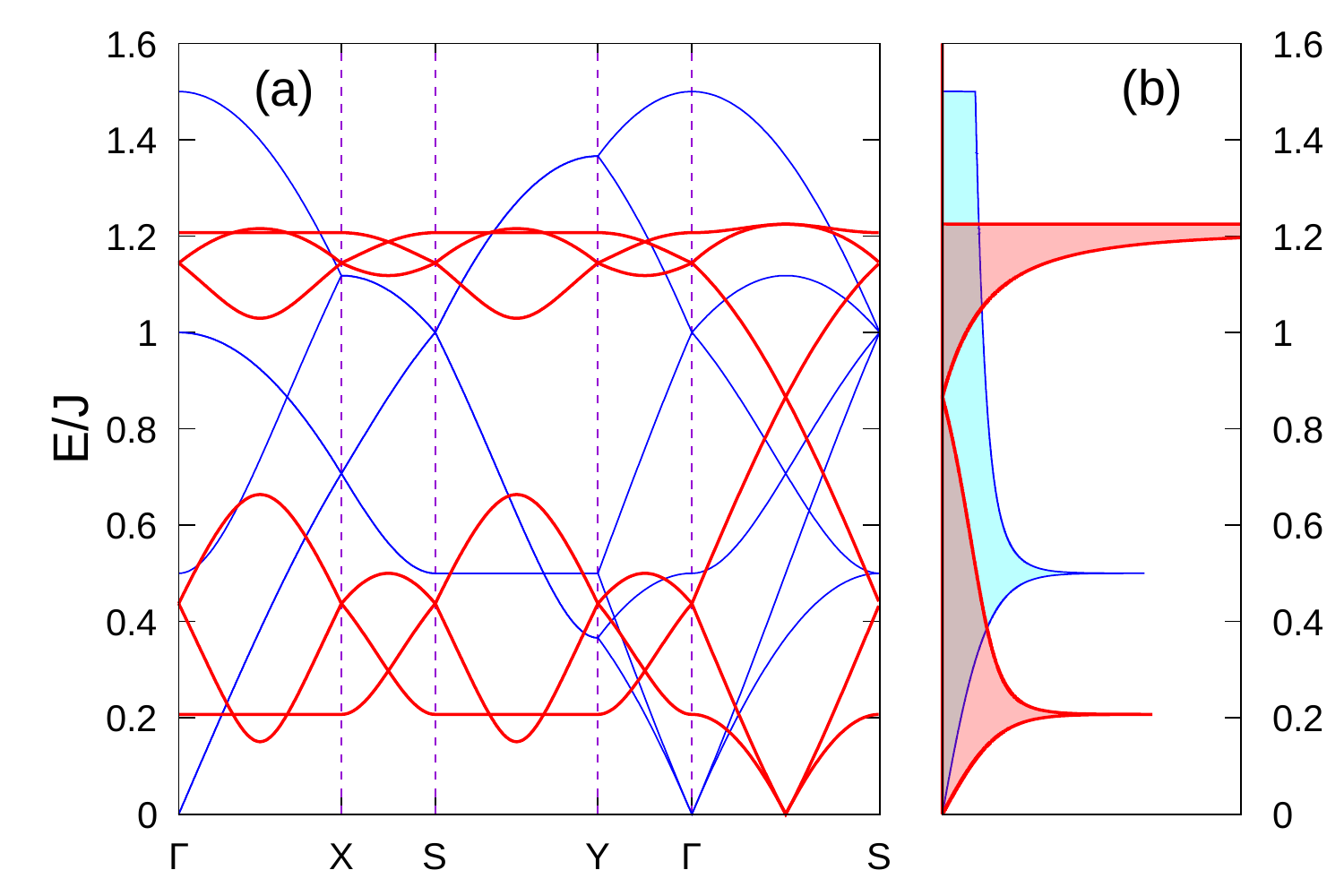}
 \caption{
   (a) Thin blue (bold red) lines represent dispersion relations
   in the system with the flux configuration I (II).
   (b) The corresponding density of states.
 }
 \label{uni}
\end{figure}

First, we focus on the uniform flux configurations I and II to discuss
the low-energy properties of the Majorana fermions and Majorana correlations in the Kitaev model.
By diagonalizing the Hamiltonians with fixed configurations
$\eta^{({\rm I})}$ and $\eta^{({\rm II})}$,
we obtain the dispersion relations and density of states,
as shown in Fig.~\ref{uni}.
It is known that, in the ground state with the flux-free configuration I,
the elementary excitations have the gapless dispersion
with the velocity $v_I=(\sqrt{3}/4)J$~\cite{Kitaev2006}.
Namely, the corresponding nodal points appear at $K$ and $K'$ points
in the original Brillouin zone, while at the $\Gamma$ point in this reduced one.
This massless dispersion leads to the long-range propagation
in the spin transport~\cite{Minakawa,Taguchi}.
We also evaluate the flux gap $\Delta_F=0.066J$~\cite{Kitaev2006,Zschocke},
which is obtained by the lowest energy change by flipping two neighboring $w_p$
in the finite cluster with $N=2\times 180\times 180$,
where $N$ is the number of sites.
It is also found that in the state with the flux configuration II,
there exist gapless excitations
and the nodal points are located between the $\Gamma$ and $S$ points,
as shown in Fig.~\ref{uni}(a).
Its velocity is given by $v_{\rm II}=(\sqrt{2}/4)J$,
which is slightly smaller than $v_{\rm I}$.
We note that, in the Majorana fermion system with the flux configuration II,
two nodal points are located at $(k_x,k_y)=(\pi/6, \sqrt{3}\pi/6)$ and
$(-\pi/6, -\sqrt{3}\pi/6)$, and
thereby the total number is four, $n_{\rm II}=4$.
This is twice larger than that for the configuration I, $n_{\rm I}=2$.
These lead to a large difference in the density of states at low-energy region
[see Fig.~\ref{uni}(b)] since it is given by $\rho(E)\sim(\sqrt{3}/4\pi)nE/v^2$.
The flux gap of the system with the flux configuration II is obtained as
$\Delta_F=0.077J$.

\begin{figure}[htb]
 \centering
 \includegraphics[width=\linewidth]{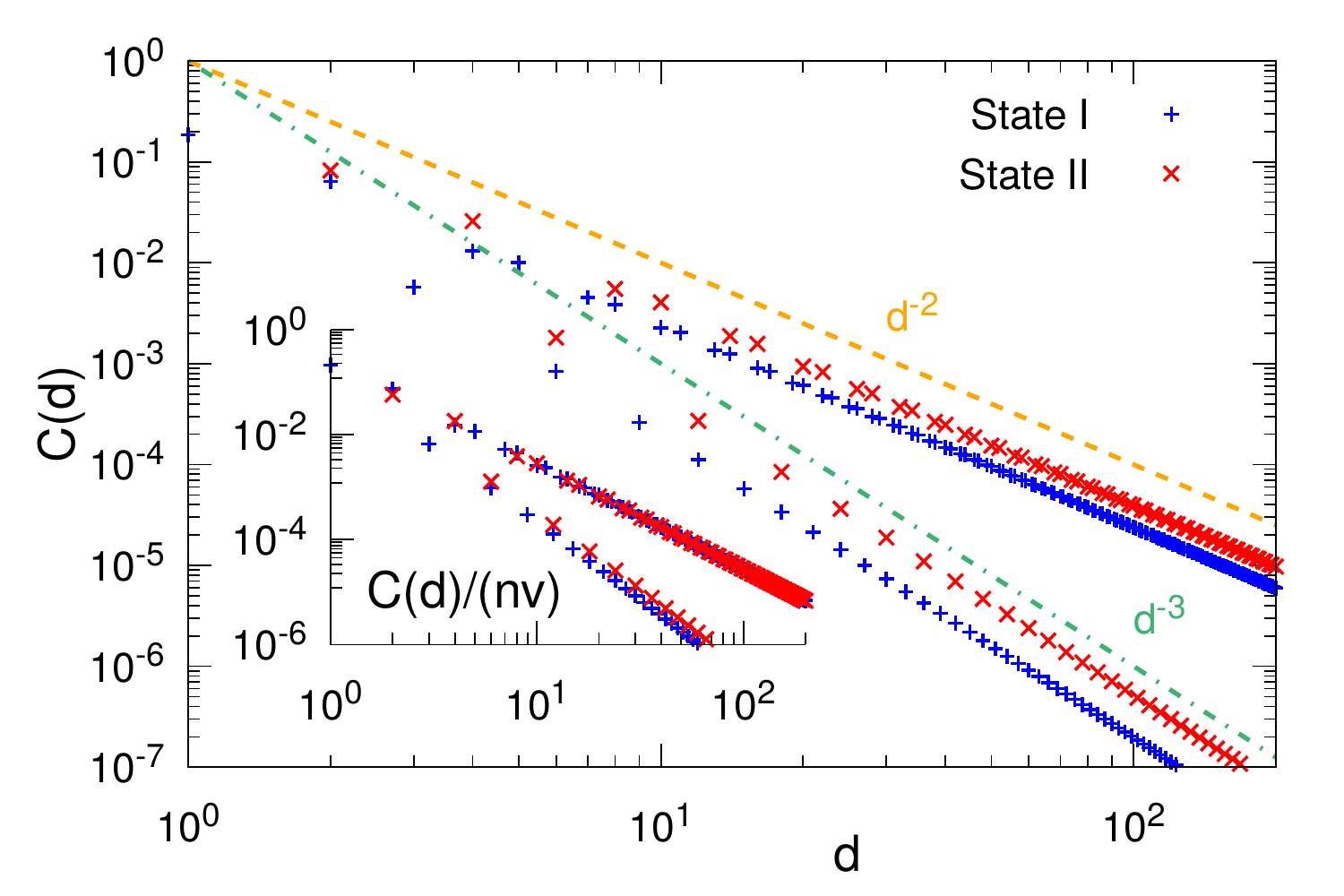}
 \caption{
   Majorana correlation $C(d)$ as a function of the distance $d$.
   Two lines with distinct powers are guides to eyes.
   The inset shows rescaled Majorana correlations 
   $C(d)/nv$.
 }
 \label{corr-uni}
\end{figure}

Figure~\ref{corr-uni} shows the Majorana correlation functions
for the ground states in two uniform flux sectors I and II along the symmetry directions.
It is found that the Majorana correlation functions decay with
period 3 in both cases.
A similar oscillatory behavior has been observed in
the finite size dependence of the energy~\cite{Kitaev2006} and flux gap~\cite{Zschocke}.
These common features originate from the Kitaev spin liquid
with the gapless linear dispersions in the Majorana fermion systems.
An important point is that both Majorana correlations exhibit two types of $d$ dependence.
One is a power-law decay with $d^{-2}$, which is a dominant contribution
while the other is scaled by $d^{-3}$.
The power law behavior is consistent with
the absence of the excitation gap in the Majorana dispersion.
This is in contrast to the fact that there are gapped spin excitations and
spin-spin correlations are extremely short-ranged.
We also find in Fig.~\ref{corr-uni} that
the Majorana correlation in the flux configuration I is smaller than that in II.
The difference comes from the number of the nodal points $n$, and their velocities $v$. 
To clarify this issue, we show $C/nv$
in the inset of Fig.~\ref{corr-uni}.
The two curves for the dominant contribution in the flux configurations I and II
appear to be on a common curve, suggesting that it is scaled by $nv$.

\begin{figure}[htb]
 \centering
 \includegraphics[width=\linewidth]{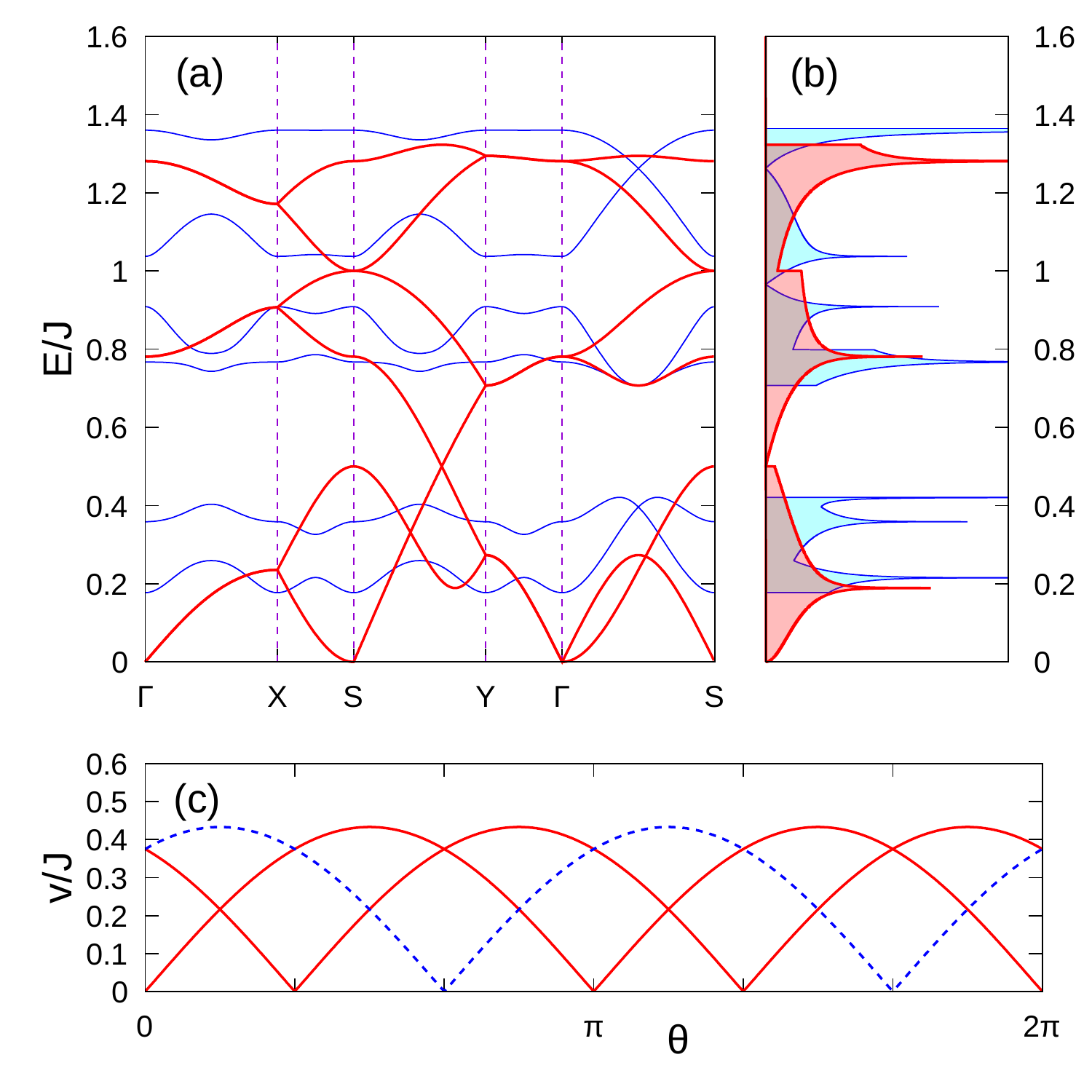}
 \caption{
   (a) Thin blue (bold red) lines represent the dispersion relations
   in the system with the flux configuration III (IV).
   (b) the corresponding density of states
   (c) Solid red (dashed blue) lines represent the Majorana velocity around
   the $\Gamma$ ($S$) point in the state with the flux configuration IV.
 }
 \label{lat}
\end{figure}

\begin{figure}[htb]
 \centering
 \includegraphics[width=\linewidth]{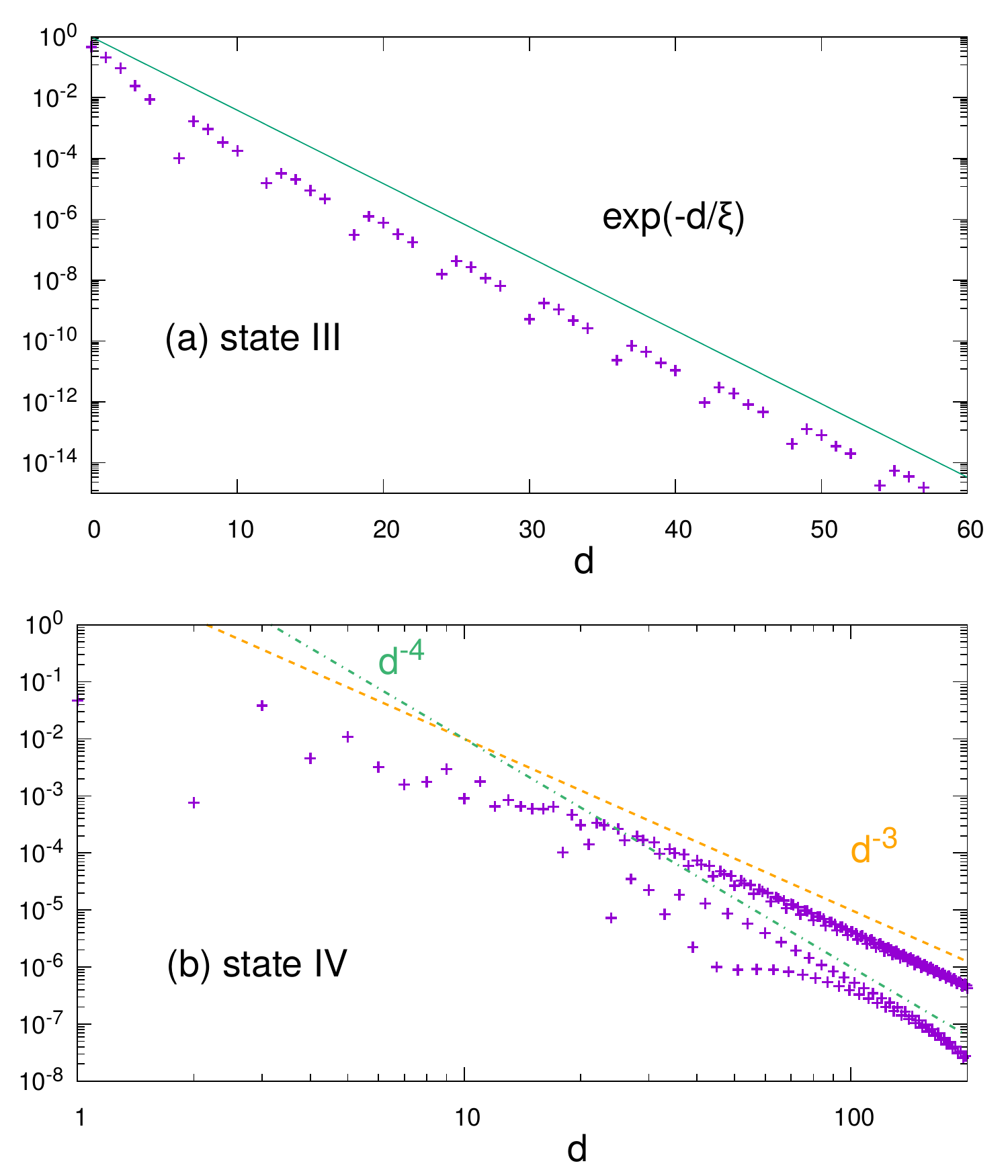}
 \caption{
   Majorana correlation functions in the system
   with the flux-lattice configurations III (a) and IV (b).
   Note that (a) and (b) are semi-log and log-log plots.
   Solid line in (a) represents the exponential function with $\xi=1.8$.
   Dashed and dot-dashed lines in (b) represent
   the functions with $d^{-3}$ and $d^{-4}$, for reference.
 }
 \label{Corr1}
\end{figure}

Next, we consider the ordered flux configurations III and IV
as shown in Figs.~\ref{lattice}(c) and \ref{lattice}(d).
These flux configurations are characterized by the same unit vectors, while
different low-energy properties appear.
The dispersion relation and density of states
are shown in Figs.~\ref{lat}(a) and \ref{lat}(b).
We find that, in Majorana fermion system with
the flux configuration III,
the finite Majorana excitation gap $(\Delta_M=0.177J)$ appears
in the low-energy region in addtion to the finite energy gap around $E=0.6J$.
This leads to the exponential decay in the Majorana correlations,
as shown in Fig.~\ref{Corr1}(a).
Therefore, we can say that the ``Majorana insulator'' is realized
by this periodically-aligned flux configuration.
Indeed, this state is topologically trivial,
which will be discussed in Appendix A.
On the other hand, the system with the flux configuration IV is gapless
at the $\Gamma$ and $S$ points, as shown in Fig.~\ref{lat}(a).
We wish to note that these points are characterized by
semi-Dirac like behavior,
where the dispersion relation is parabolic
along a certain axis and linear along the other.
Namely, there are two semi-Dirac like dispersions at the $\Gamma$ point
and one at the $S$ point.
Figure~\ref{lat}(c) shows the angle $\theta$ dependence of the velocity
at ${\bf k}_0(=\Gamma$ or $S$),
where it is defined by $v(\theta) =\lim_{\Delta {\bf k}\rightarrow 0}
E({\bf k}_0+\Delta{\bf k})/|\Delta{\bf k}|$
with $\theta=\tan^{-1}\Delta k_y/\Delta k_x$.
It is found that, in the case with $[\theta=m\pi/3\;(m=0,1,\cdots, 5)]$,
one of three velocities vanishes and
its dispersion is parabolic.
This yields interesting low-energy dependence in the density of states,
$\rho(E)\sim E^{1/2}$.
Figure~\ref{Corr1}(b) shows that Majorana correlations
in the state with the flux configuration IV.
We find that, in the short range region $(d<40)$,
the values of correlations are randomly distributed,
in contrast to the configurations I, II, and III with a certain periodicity in $C(d)$.
On the other hand, power law behavior clearly appears in larger $d$ region,
where Majorana correlation function obeys $d^{-3}$ dominantly
and smaller correlations show a decay with $d^{-4}$.
The behavior is in contrast to that in the cases I and II with
the isotropic linear dispersion discussed above.
Namely, Majorana correlations in the other directions should obey
$d^{-2}$ since Majorana fermions have
the finite velocity, as shown in Fig.~\ref{lat}(c).

\subsection{Effects of the disorder in flux configurations}
\begin{figure}[htb]
 \centering
 \includegraphics[width=\linewidth]{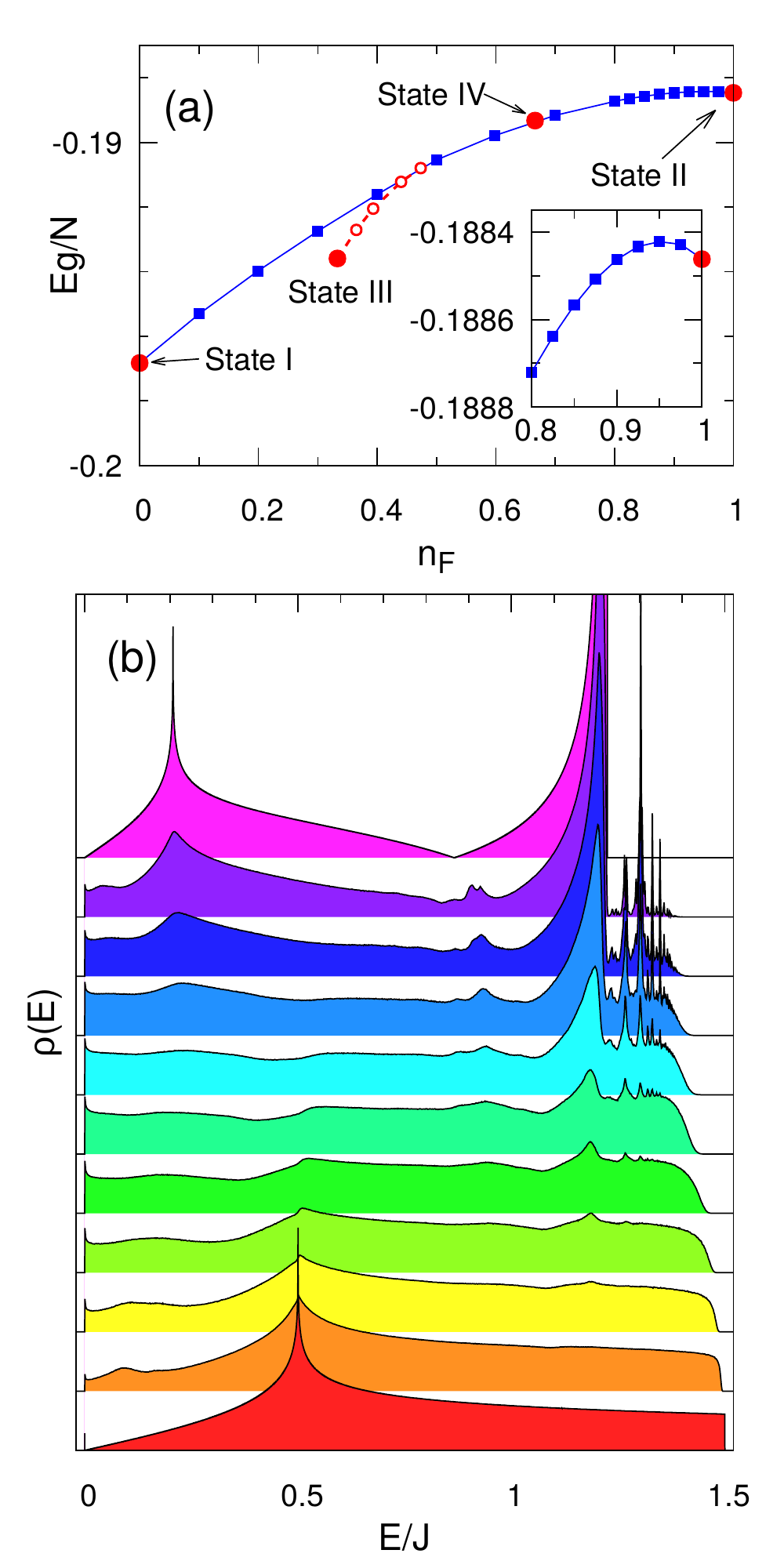}
 \caption{
   (a)~Ground state energy as a function of the flux density
   in the system with the random configurations $(N=2\times180\times 180)$.
   Open circles with the dashed line represent the results for the effect of
   the flux disorder in the configuration III (see text).
   (b)~Density of states in the Kitaev model
   with disordered configurations.
   The data are for $n_F=0, 0.1, 0.2, \cdots$, and $1$ from the bottom to the top.
 }
 \label{random}
\end{figure}

We also consider the effects of the flux disorder in the ordered-flux Kitaev systems
discussed above,
which should be important to understand how robust
the spin transport is against thermal fluctuations
in the realistic materials.
Now, the flux density is defined as $n_F=N_F/N_p=2N_F/N$, 
where $N_F$ and $N_p$ are the numbers of fluxes and plaquettes,
respectively.
In the system with $n_F=0.5$, the fluxes are randomly distributed,
and its ground state should capture the essence of the Kitaev system at
the intermediate temperatures~\cite{Kao2020,Kao2021}.
When $n_F=0$ $(n_F=1)$, the flux-free (full-flux) configuration is realized
with long-range Majorana correlations.
By contrast, in the general case with $0<n_F<1$, 
one may expect that the randomness yields the localization
in each wave function, leading to short-ranged correlations.
Therefore, it is not clear how the Majorana correlations
are changed by the flux disorder.
To clarify the effects of the random flux configurations,
we prepare more than hundred distinct clusters with $N=2\times 180 \times 180$.
We diagonalize the Hamiltonian for each flux configuration  
by means of the singular value decomposition~\cite{Pedrocchi,Zschocke}.
The obtained ground state energy $E_g$ as a function of the flux density $n_F$
is shown in Fig.~\ref{random}(a).
We find that the minimum of the curve is located at $n_F=0$,
which is consistent with the fact that
the ground state of the Kitaev model is indeed realized
in the flux-free sector~\cite{Kitaev2006}.
The maximum is located around $n_F=0.95$.
This implies that the finite energy is necessary to remove a flux
in the state with the configuration II, as discussed above.
Due to the convex structure in the energy curve,
one may expect that the phase separation occurs in the Kitaev system
with the fixed flux density.
However, $W_p$ is a local conserved quantity in the Kitaev model Eq.~(\ref{eq:Model}),
and thereby its configuration is never changed.
Therefore, the phase separation does not occur
in the disordered state.

Figure~\ref{random}(b) shows the density of states for the system with
random configurations.
When $n_F=0$, linear behavior appears around $E\sim 0$ and
the peak structure appears around $E/J=0.5$.
Introducing the fluxes, we find that the peak structure smears
and the density of states around lower energy states increases.
In particular, a sharp peak structure develops
around $E\sim 0$,
suggesting that low-energy excitations are induced by the random fluxes.
Similar behavior also appears when some fluxes with $w_p=-1$ are randomly
inverted to $w_p=+1$
in the state with the flux configuration II ($n_F=1$). 
Namely, the lowest energy level is evaluated
around $\Delta_M/J \sim 10^{-5}$
in the finite cluster with $0<n_F<1$,
which should indicate that the disordered system
is gapless in the thermodynamic limit.

\begin{figure}[htb]
 \centering
 \includegraphics[width=\linewidth]{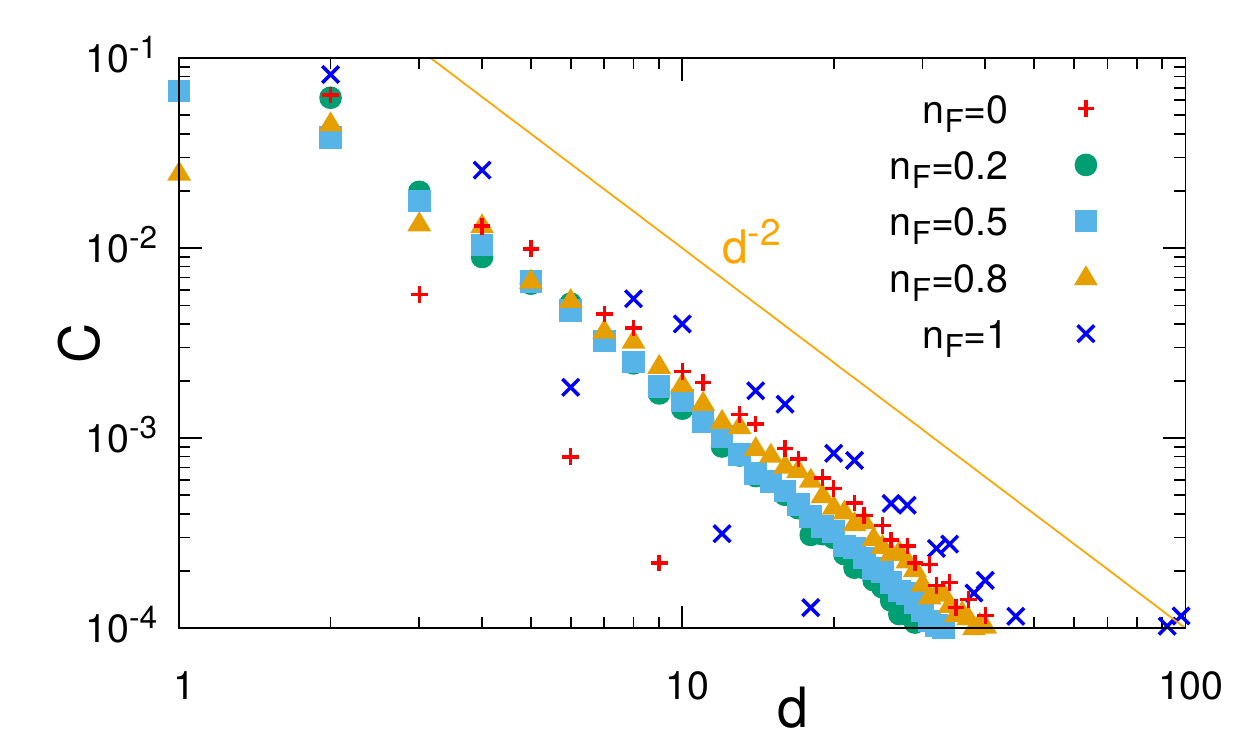}
 \caption{
   Majorana correlation function in the Kitaev model
   with disordered configurations.
   The data are averaged over more than 100 samples
   for the cluster with $N=180\times~180\times~2$.
  }
 \label{randomcorr}
\end{figure}

Although low-energy properties are clarified in the disordered systems,
it is still unclear whether Majorana correlations
are long-ranged or not.
To clarify this,
we calculate the Majorana correlations.
The obtained results are shown in Fig.~\ref{randomcorr}.
Our results for each $n_F$ suggest that
the correlation function obeys the power law 
with respect to distance
although the boundary effect appears $d\gtrsim 30$.
This is similar to the results for both uniform flux configurations I and II.
Therefore, the randomness in the flux configuration
little affects low-energy properties of the Kitaev spin liquid,
which is consistent with the recent results~\cite{Kao2021}.

\begin{figure}[htb]
 \centering
 \includegraphics[width=\linewidth]{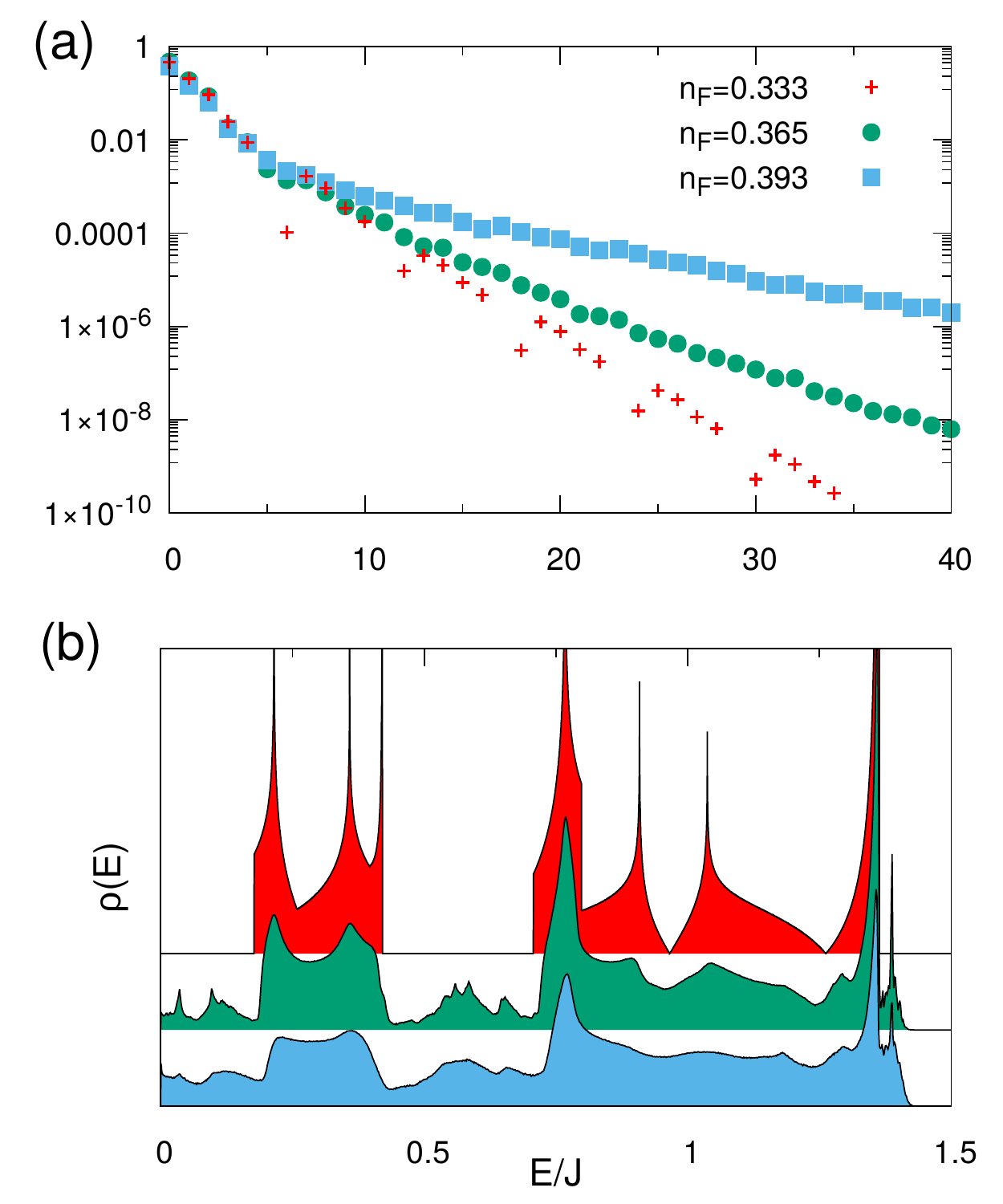}
 \caption{
   (a) Majorana correlation function and (b) density of states
   in the Kitaev model.
   The flux configurations for the flux densities $n_F=0.333$, $0.365$, and $0.393$
   are generated by flipping $\{\eta_r\}$ in the configuration III
   with the probabilities $p_3=0, 0.05$, and $0.1$, respectively.
   The data are averaged over more than 100 samples
   for the cluster with $N=180\times~180\times~2$.
  }
 \label{randomcorr2}
\end{figure}

We also discuss the ``Majorana insulator'' with the flux configuration III.
Its energy $E_g/N=-0.194$ is lower than the random flux states with $n_F=1/3$
($E_g/N=-0.192$),
which should originate from the gap formation in the Majorana dispersion
[see Fig.~\ref{random}(a)].
To examine the stability of the state against the flux disorder,
we prepare the flux disordered configurations, which are obtained by
flipping $\{\eta_r\}$ in the configuration III
with a certain probability $p_3$.
The energy for this disordered system is shown as
the open circles with the dashed line in Fig.~\ref{random}.
By introducing the disordered fluxes in the configuration III,
the properties of the Majorana fermions inherent 
in the configuration III smear.
In fact, the energy increases and approaches the energy curve
obtained above around $n_F\sim 0.4$.
As for Majorana correlations, power law behavior appear
if one focuses on long range behavior.
Nevertheless, the rapid decrease still appears in the short range correlations $(d\lesssim 5)$,
as shown in Fig.~\ref{randomcorr2}(a).
This implies that the Kitaev system with the gapped flux configuration III
is stable against the flux disorder.
This is consistent with the fact that the gap structure in the density of states
still remains although low-energy states are induced by the disorder, 
as shown in Fig.~\ref{randomcorr2}(b).

Before conclusion, we would like to comment on the spin transport
in the Kitaev system although it is beyond the scope of the present study.
It has been clarified that, in the Kitaev model,
the spin transport is mediated by
the itinerant Majorana fermions~\cite{Minakawa}.
This phenomenon is dominated by the velocity of 
gapless Majorana fermions,
which is regarded as the ``Majorana metal''.
Owing to this feature in the Majorana fermion system, long-range spin transport is realized despite the presence of the spin gap.
On the other hand, the flux configuration III yields the excitation gap
in the Majorana dispersion, 
corresponding to the ``Majorana insulator'', where the spin transport shows an exponential decay.
Exploiting these features, 
one could manipulate the motion of carriers of the spin excitations on the basis of
 the flux configuration, which may open the Majorana-mediated spintronics.
It is an interesting problem to clarify how the flux configurations are
controlled in the realistic materials, which is now under consideration.

\section{Summary}
We have studied the $S=1/2$ Kitaev model on the honeycomb lattice
to reveal how flux configurations affect Majorana correlations.
It has been clarified that the systems with the uniform flux configurations
have linear dispersions with nodal points and  
a power law behavior with $d^{-2}$ appears in the Majorana correlations.
The system with ordered flux configuration III has the gapped dispersion and
the exponential decay appears in the Majorana correlations.
This means that Majorana insulator is realized
in terms of this flux configuration.
On the other hand, the Kitaev system with the configuration IV exhibits
the semi-Dirac like dispersion,
leading to the power law decay with $d^{-3}$.
We have discussed the effect of the randomness in the flux configuration
to clarify that power law behavior appears in the Majorana correlations.
It is also interesting to discuss how robust Majorana correlations are
in the related models such as
the bilayer Kitaev model~\cite{Tomishige1,Seifert,Tomishige2},
Kitaev-Heisenberg model~\cite{Chaloupka_2010}, and
higher spin models~\cite{Baskaran2007,S1Koga,Oitmaa,S1Trans,Stavropoulos_2019,Dong_2020,Lee_2020}.

\begin{acknowledgments}
  We would like to thank T. Minakawa for valuable discussions.
  Parts of the numerical calculations are performed
  in the supercomputing systems in ISSP, the University of Tokyo.
  This work was supported by Grant-in-Aid for Scientific Research from
  JSPS, KAKENHI Grant Nos.
  JP19H05821, JP18K04678, JP17K05536 (A.K.),
  JP19K23425, JP20K14412, JP20H05265 (Y.M.), JP19K03742 (J.N.), 
  JST CREST Grant No. JPMJCR1901 (Y.M.), and JST PREST No. JPMJPR19L5 (J.N.).

\end{acknowledgments}

\appendix
\section{The effect of the three-spin interaction}
\begin{figure}[htb]
 \centering
 \includegraphics[width=0.8\linewidth]{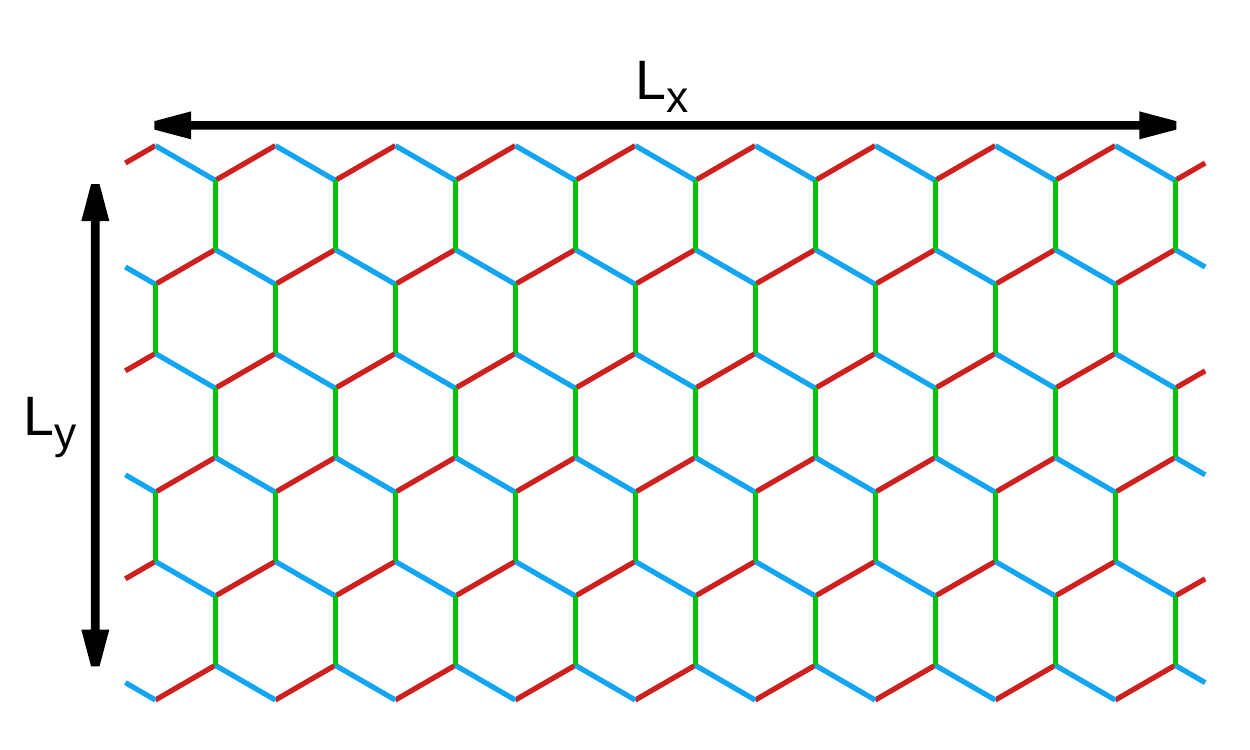}
 \caption{
   Lattice structure of the Kitaev model with zigzag edges.
 }
 \label{ribbon}
\end{figure}

In the appendix, we consider the three-spin interactions,
which break the time-reversal symmetry,
to clarify how the topological state is realized~\cite{Kitaev2006,Nasu2,Nasu3,Hickey_2019}.
We also clarify that the gapped state with the flux configuration III
is topologically trivial.
Here, we introduce the interaction term defined by
\begin{eqnarray}
  H'_{\rm eff}&=&h\sum_{(ijk)}S_i^\alpha S_j^\beta S_k^\gamma,\label{3}
\end{eqnarray}
where $h$ is the magnitude of interaction and
the summation takes 
 over adjacent three spins:
{\it eg.} $(i,j,k)=(p_2,p_3,p_4)$ and $(\alpha,\beta,\gamma)=(x,y,z)$.
The other symmetry-equivalent pairs are shown in Fig.~\ref{simple}(b).
We note that this 
 is derived as the low-energy Hamiltonian
for the ground state
by means of the third-order perturbation theory for the Zeeman Hamiltonian
${\cal H}'=-\sum_i \left(h'_x S_i^x+h'_yS_i^y+h'_zS_i^z\right)$,
where ${\bf h}'$ is the external magnetic field~\cite{Kitaev2006}.
The Hamiltonian Eq.~(\ref{3}) is represented in terms of the Majorana operators as,
\begin{eqnarray}
  H'_{\rm eff}
  &=&-\frac{ih}{8}\sum_p \left(
    \gamma_{p1}\gamma_{p5}+\eta_{pr}\gamma_{p1}\gamma_{p5}+\eta_{pl}\gamma_{p3}\gamma_{p1}\right.\nonumber\\
    &&\left.
    +\gamma_{p4}\gamma_{p2}+\eta_{pl}\gamma_{p2}\gamma_{p6}+\eta_{pr}\gamma_{p6}\gamma_{p4}
    \right).\label{Zeeman}
\end{eqnarray}
Since this Hamiltonian is quadratic, one discusses low-energy properties
in the same framework as Eq.~\eqref{eq:Majorana}.
Here,
we consider the Kitaev model with the edges, as shown in Fig.~\ref{ribbon}.
This allows us to examine how three-spin interactions induce
the excitation gap, and
whether or not the topological edge modes
are induced inside of the excitation gap.

\begin{figure}[htb]
 \centering
 \includegraphics[width=\linewidth]{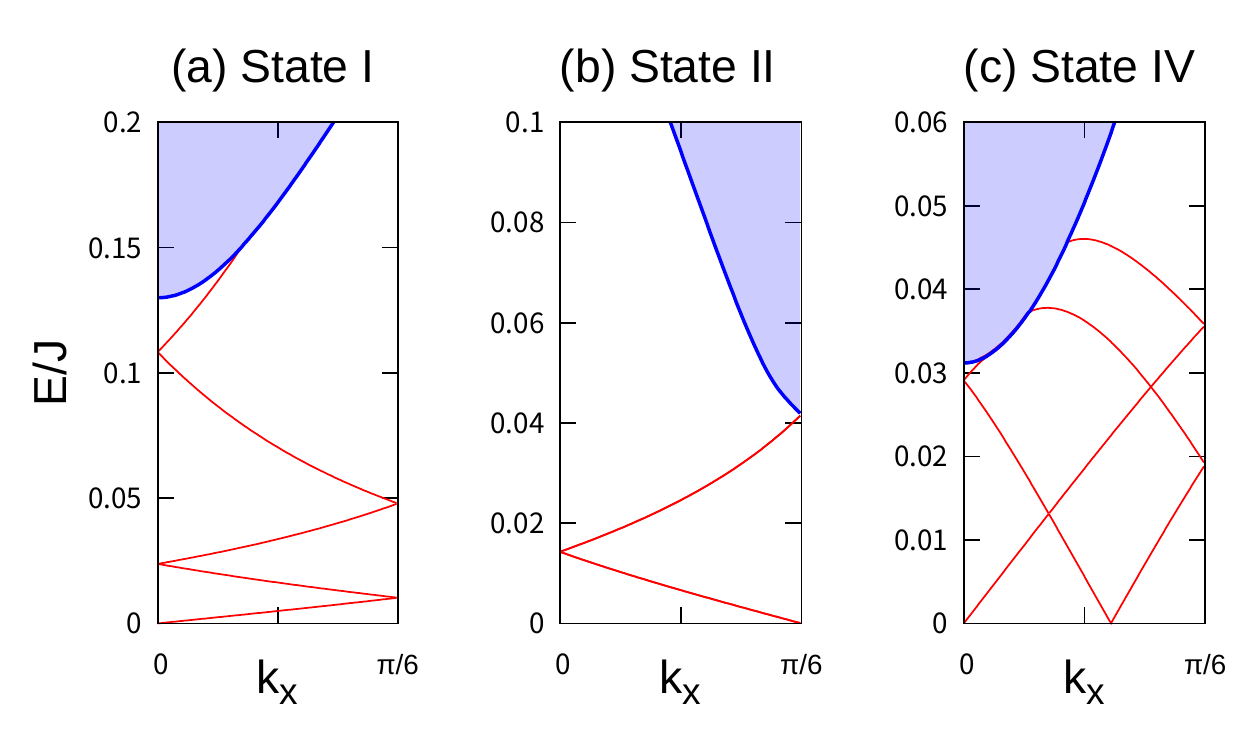}
 \caption{
   Dispersion relations in the zigzag-edge Kitaev model with
   flux configurations I (a), II (b), and IV (c) when
   $h/J=0.1, 0.1$, and $0.4$, respectively.
   The shaded regions represent the continuum for the bulk dispersion.
 }
 \label{com023}
\end{figure}
First, we deal with the gapless systems with flux configurations I, II, and IV.
Figure~\ref{com023} shows the dispersion relations in the Kitaev model
with zigzag edges.
In the system with the configuration I,
the interaction induces the excitation gap with
$\Delta=(3\sqrt{3}/4)h$
and edge modes inside the gap~\cite{Kitaev2006,RevJPSJ}.
Similar behavior appears in the system with the configuration II,
where the bulk excitation gap $\Delta=(\sqrt{3}/4) h$ appears with topological edge states.
As for the semi-Dirac like system with the configuration IV,
we find that the excitation gap is induced with $\Delta=(7/32) h^2$
and some topological edge modes appear.
Therefore, we can say that the three-spin terms drives the gapless systems
to the topological state with a finite bulk gap.

\begin{figure}[htb]
 \centering
 \includegraphics[width=\linewidth]{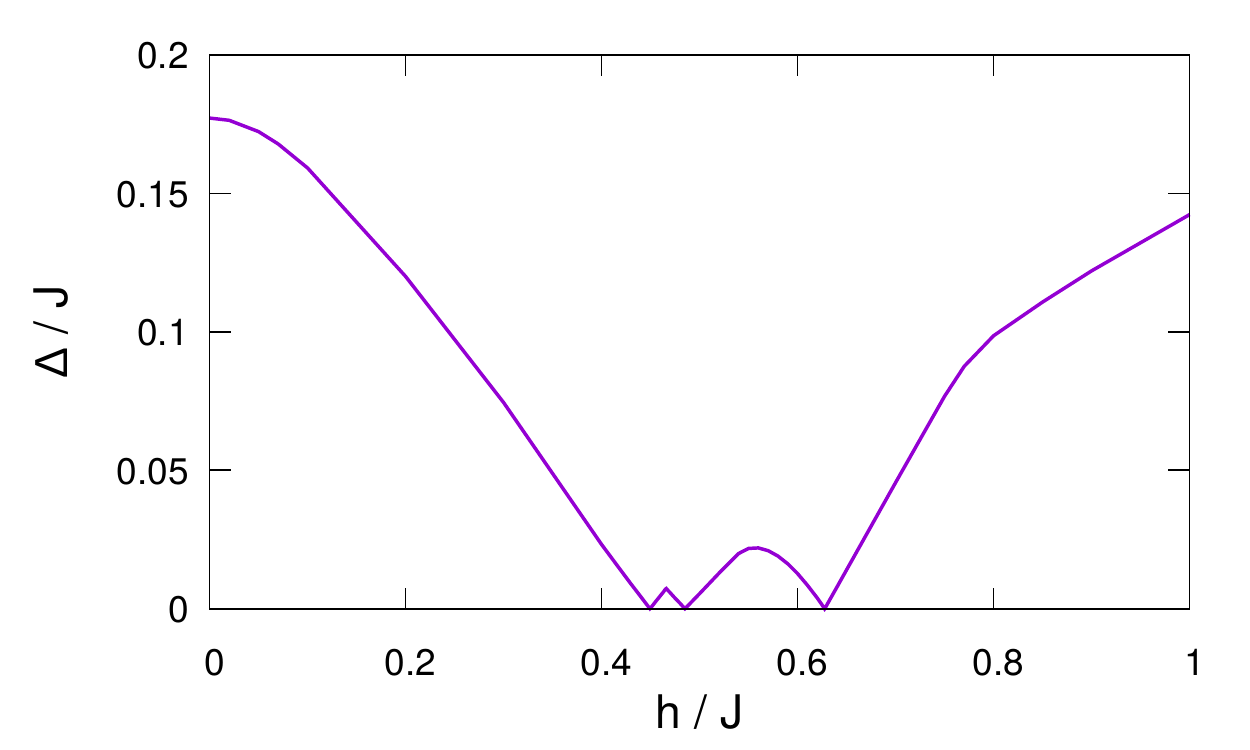}
 \caption{
   Majorana excitation gap 
   $\Delta$ as a function of $h$
   in the Kitaev model with the flux-lattice configuration III.
 }
 \label{gapIII}
\end{figure}

\begin{figure}[htb]
 \centering
 \includegraphics[width=\linewidth]{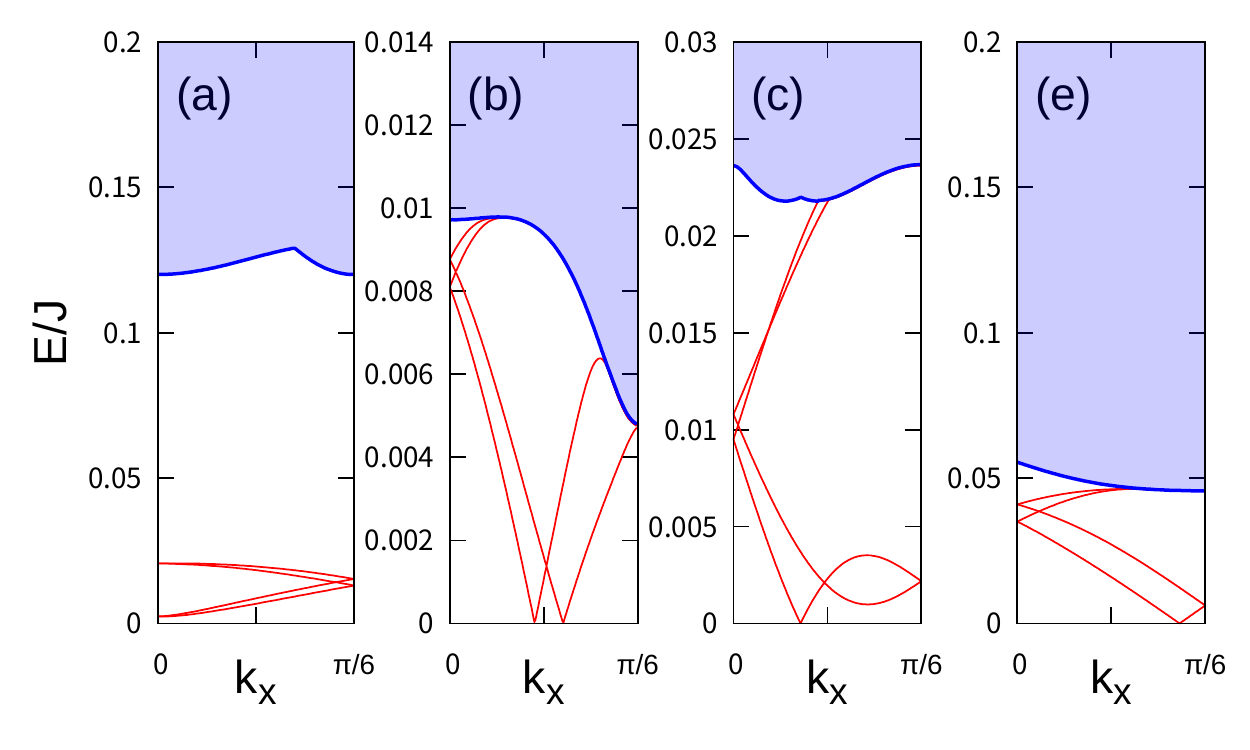}
 \caption{
   Dispersion relations in the zigzag-edge Kitaev model with
   flux configuration III when
   $h/J=0.2$ (a), $0.46$ (b), $0.55$ (c), and $0.8$ (d).
   The shaded regions represent the continuum for the bulk dispersion.
 }
 \label{com1}
\end{figure}

As discussed in the text, the system with the flux configuration III
has the gap in the Majorana excitation.
Now, we consider how stable the gapped state is against
the three-spin interactions.
Figure~\ref{gapIII} shows the Majorana excitation gap in the bulk system. 
It is found that, by introducing the interactions,
the excitation gap decreases and finally reaches zero at
the critical value $(h/J)_{c1}\sim 0.45$.
Beyond the critical value,
we find two additional critical values $(h/J)_{c2}\sim 0.485$ and
$(h/J)_{c3}\sim 0.628$. 
These mean the existence of, at least, three phase transitions.
To clarify the topological nature of four distinct phases,
we show in Fig.~\ref{com1} the dispersion relations
in the system with zigzag edges.
When $h/J=0.2$, the edge modes are below the bulk continuum.
Therefore, this gapped state is topologically trivial.
On the other hand, when $h_{c1}<h$,
there exist topological edge states in the inside of the bulk gap.
Therefore, topological phase transitions occur in the system with the flux configuration III.
Beyond the critical value $h_{c1}$,
the number of edge states is finite and topological states are realized.

\bibliography{./refs}

\end{document}